\documentclass[aps,prl,twocolumn,noshowpacs]{revtex4-1}
\usepackage{amsmath}
\usepackage{color}
\usepackage{graphicx}
\usepackage{epstopdf}

\begin{document}

\title{Lattice matched Volmer-Weber growth of Fe$_3$Si on GaAs(001) -- the influence of the growth rate}%

\author{B. Jenichen}
\email{bernd.jenichen@pdi-berlin.de}
\author{Z. Cheng}
\author{M. Hanke}
\author{J. Herfort}
\author{A. Trampert}

\affiliation{Paul-Drude-Institut f\"{u}r Festk\"{o}rperelektronik Leibniz-Institut im Forschungsverbund Berlin e.V., Hausvogteiplatz 5-7, 10117 Berlin, Germany}%

\date{\today }

\begin{abstract}
We investigate the formation of lattice matched single-crystalline Fe$_3$Si/GaAs(001) ferromagnet/semiconductor hybrid structures by Volmer-Weber island growth, starting from the epitaxial growth of isolated Fe$_3$Si islands up to the formation of continuous films as a result of island coalescence. We find coherent defect-free layers exhibiting compositional disorder near the Fe$_3$Si/GaAs--interface for higher growth rates, whereas they are fully ordered for lower growth rates.
\end{abstract}

\maketitle

\section{Introduction}

Often thin film growth can be realized in a layer-by-layer mode resulting in narrow interface widths. However, during heteroepitaxy of a metal film on top of a semiconductor different surface tensions of the epitaxial materials play a crucial role.\cite{Volmer1926,Ravisvaran2001} A poor wetting of the substrate surface by the deposited metal may result in an island growth mode even for zero mismatch,\cite{kag09} which can be utilized for the growth of nanostructures.\cite{Placidi2000,Ravisvaran2001} Surface energies of GaAs(001) lie in the range near 65~$meV/{\AA}^2$ \cite{moll1996} whereas the energies of Fe$_3$Si(001) are in the range from 100 to 200~$meV/{\AA}^2$ or even above,\cite{Hafner2007} i.e. Fe$_3$Si has really higher surface energies compared to GaAs.

\begin{figure}[h]
\includegraphics[width=0.6\linewidth,angle=0, clip]{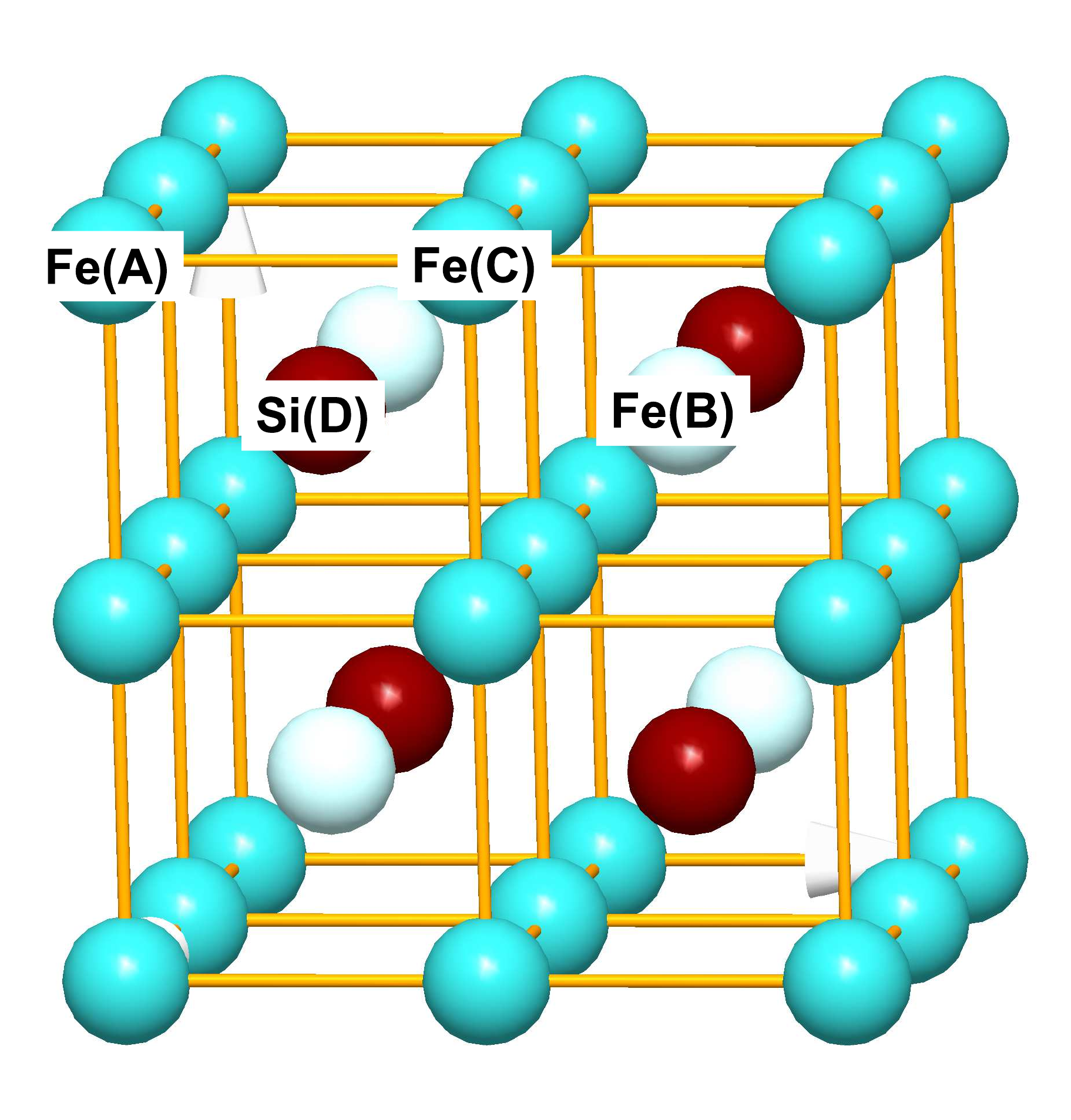}
\caption {D0$_3$ order of the Fe$_3$Si lattice. The Si-atoms are located on the D-position of the lattice, whereas the Fe-atoms sit on the A-, B-, and C-positions.}
\label{order}
\end{figure}

In the case of Fe$_3$Si growth on GaAs the stoichiometry of the metallic films has an influence on their lattice parameter\cite{herfort03,Herfort2004} and their long-range ordering.\cite{Jenichen05}
We find three types of diffraction maxima of Fe$_3$Si. Fundamental reflections, i.e. $H~+~K~+~L~=~4n$ (where $n$ is integer),  are not sensitive to disorder. Their structure factor is $F_{4n} = 4(f_{Si}+3f_{Fe})$. In the D$0_3$  structure of Fe$_3$Si the Si atoms occupy the lattice position D, whereas the Fe atoms sit on the positions A,B,C, see Figure~\ref{order}.\cite{niculescu76,Jenichen05}  The disorder is described by two types of order parameters $\alpha$ and $\beta$, which are the fractions of Si atoms occupying the Fe(B) and the Fe(A,C) sites, respectively. For $H~+~K~+~L~=~2n$ (where $n$ is odd) the structure factor is
\begin{equation}
F_{2n} = -4(1-2\beta)(f_{Si}-f_{Fe}).
\end{equation}
\label{fundamental}

And for odd H,K,L we have
\begin{equation}
F_{2n+1} = 4i(1-2\alpha-\beta)(f_{Si}-f_{Fe}).
\end{equation}

The lattice misfit between Fe$_3$Si and GaAs is minimized for stoichiometric films. The lattice parameter of GaAs is 0.56325~nm whereas the lattice parameter of Fe$_3$Si is 0.5654~nm.\cite{Hongzhi2007} From these values we obtain a mismatch below 0.4~\%. In earlier work we found evidence for the presence of islands: The measured island height was larger than the nominal thickness of the deposited film.\cite{kag09} Later Fe$_3$Si islands on GaAs were directly imaged by scanning tunneling microscopy, and it was found that the islands show a D0$_3$ structure, i.e. they were fully ordered.\cite{noor2013} After coalescence of the Fe$_3$Si islands a layer-by-layer growth of the metal was observed, which is typical for homoepitaxy of Fe$_3$Si.\cite{jen07, jen08a}

In general, the material Fe$_3$Si with its high Curie temperature of about 567$^{\circ}$C is well suited for spintronic applications. The spin polarization of Fe$_3$Si is about 45\%.\cite{ionescu05} Room temperature spin injection from Fe$_3$Si into GaAs was demonstrated.\cite{Herfort2005} The role of interdiffusion in the system Fe$_3$Si/GaAs was investigated, and influence of interdiffusion on the ordering was found.\cite{Gusenbauer2011,Krumme2009} The ferromagnetism of the thin Fe$_3$Si films arises at a nominal thickness of about 3~monolayers (MLs).\cite{liou93,Herfort2008,noor2013} One ML corresponds to 0.28~nm.

The aim of the present work is a detailed structural characterization of the heteroepitaxial Fe$_3$Si on GaAs(001). We directly image Fe$_3$Si growth islands by cross-section high resolution transmission electron microscopy (HR TEM) and perform corresponding measurements of crystal truncation rods using grazing incidence X-ray diffraction of synchrotron radiation. The influence of the growth rate on long-range ordering is studied. A comparison of fundamental and superlattice maxima gives information about long-range ordering within the Fe$_3$Si.\cite{niculescu76,Jenichen05} Residual disordering near the Fe$_3$Si/GaAs interface is revealed using the Z-contrast method in a  probe-C$_s$-corrected scanning TEM with atomic resolution.


\begin{table*}[htbp]
  \caption{Nominal (measured) film thicknesses (island heights), substrate temperatures T$_S$, growth rates v$_g$  during epitaxial growth, and the order parameters $\beta$ determined by simulation of the X-ray diffraction L-scans for four samples investigated.}

    \begin{tabular}{l c c c c c c c c c c c c c c c c}
    \toprule
          & sample & 1 &  &  & sample & 2 &  &  & sample & 3 & &  & sample & 4 &  & \\
      & thickness & T$_S$ & v$_g$ & $\beta$ & thickness &  T$_S$ &  v$_g$ & $\beta$ & thickness & T$_S$  & v$_g$ & $\beta$ & thickness &  T$_S$ &  v$_g$ & $\beta$  \\

          & (ML)  & $^\circ$C & (ML/h) &  & (ML) & $^\circ$C & (ML/h) &  & (ML) & $^\circ$C &  (ML/h) &  & (ML) & $^\circ$C & (ML/h) &   \\
    \hline
    \multicolumn{1}{c}{GaAs} & 1071     & 580 &  & ./. & 1071 &   580 &  & ./. & 1071 & 580 & & ./. & 1071 &   580 &  & ./. \\
    \multicolumn{1}{c}{Fe$_{3}$Si} & 3 (4) & 200 & 71 & 0.48   & 3 (4) & 200 & 3 & 0.0  & 6 (7) & 200 & 71 &0.45 & 7 (7$\pm$1) & 200 & 3 & 0.0   \\
    \multicolumn{1}{c}{Ge} & 14 & 150 &   & ./.  & ./. &   &   & ./.   & 14 & 150 & & ./. & ./. &   &   & ./.   \\
   \toprule
    \end{tabular}%
  \label{tab:tab1}%
\end{table*}%

\begin{figure}[h]
\includegraphics[width=0.8\linewidth,angle=0, clip]{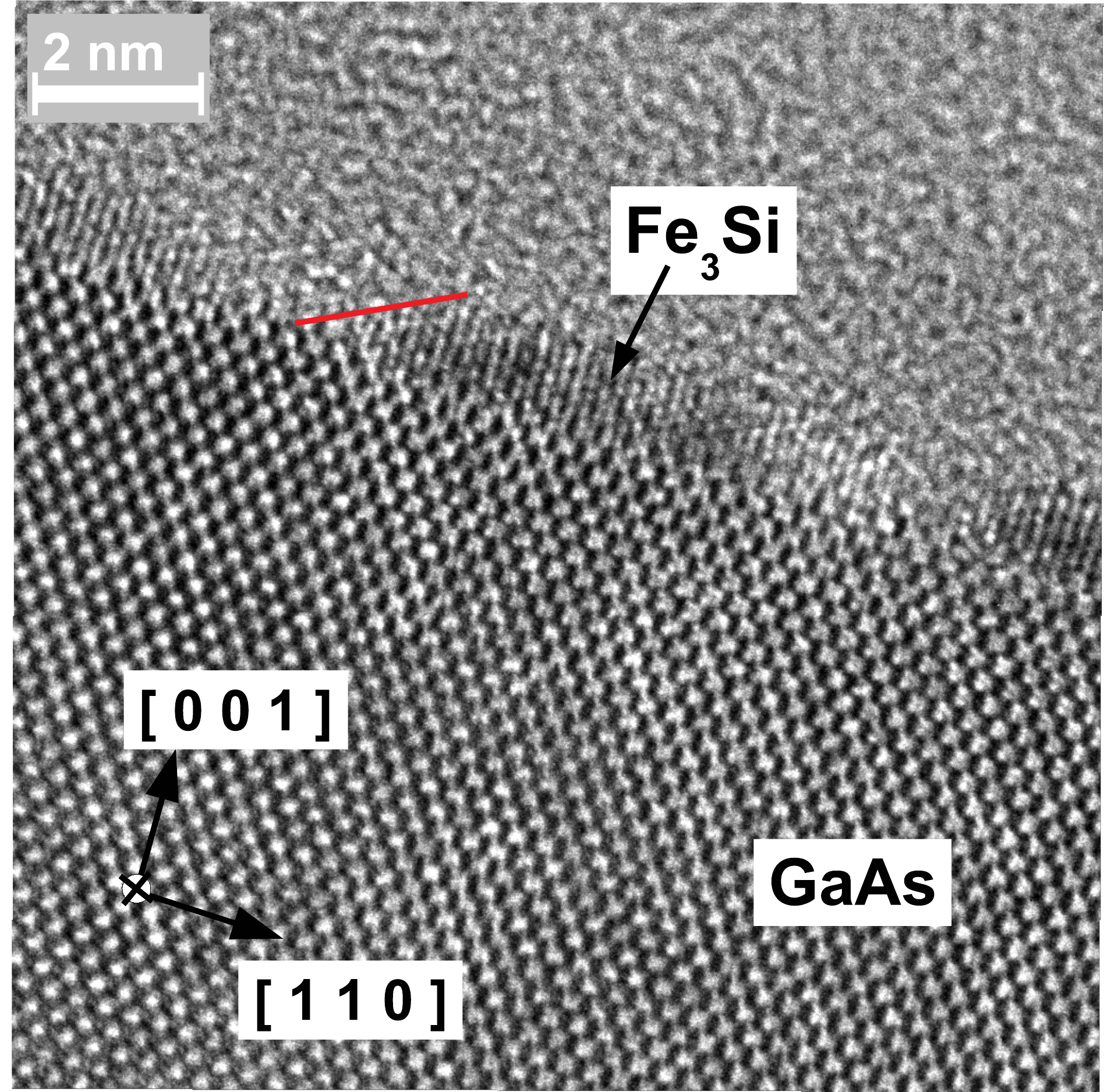}
\caption {Cross-section high-resolution transmission electron micrograph of sample~1 with Fe$_3$Si islands epitaxially grown on the GaAs(001) substrate. The contact angle at the edge of the Fe$_3$Si island is marked by a red line. }
\label{islands}
\end{figure}

\begin{figure}[h]
\includegraphics[width=1.1\linewidth,angle=0, clip]{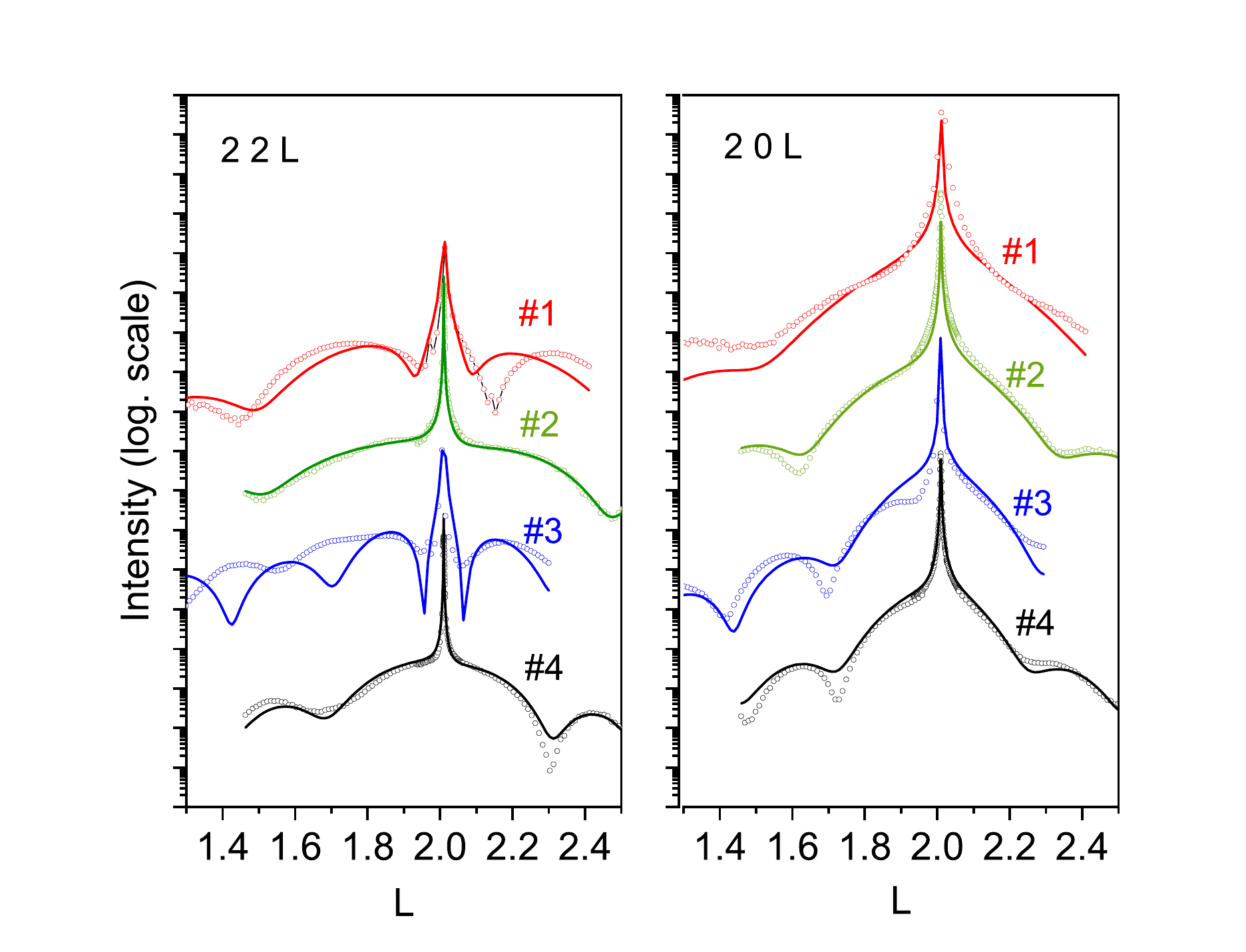}

\caption {X-ray diffraction L-scans along the 22L (left) and 20L (right) crystal truncation rods for all samples. The symbols show the results of the experiments, the lines are the corresponding simulations. For sample~1 the intensity of the Fe$_3$Si 222 maximum is reduced near the GaAs 222 peak. For sample~2 we observe the full intensity of the Fe$_3$Si 222 maximum. For sample~3 the intensity of the Fe$_3$Si 222 maximum is reduced near the GaAs 222 peak, and the disorder is changing with depth. For sample~4 we observe again the full intensity of the Fe$_3$Si 222 maximum. The resulting order parameters and film thicknesses are given in Table~\ref{tab:tab1}.}
\label{XRD}
\end{figure}

\section{Experimental}
The GaAs(001) substrates were overgrown with a 350~nm thick GaAs buffer layer at a growth temperature T$_G$ = 580$^{\circ}$C. After cooling down this leads to a formation of an atomically flat and As-rich c(4$\times$4) reconstructed GaAs(001) surface. Subsequently, the substrates were transferred under UHV conditions to a separate, As-free chamber with a base pressure of 1$\times$10$^{-10}$~mbar where the Fe$_3$Si was grown at different growth rates (3~ML/h and 71~ML/h). Fe and Si where coevaporated and deposited on the GaAs substrate at T$_G$ = 200$^{\circ}$C.\cite{herfort03,Herfort2004}

Two types of samples were compared (see Table~\ref{tab:tab1}), i.e. samples grown with a relatively high Fe$_3$Si growth rate (samples~1~and~3) and samples grown with lower Fe$_3$Si growth rate (samples~2~and~4). Growth rates and Fe$_3$Si stoichiometry were determined via calibration measurements using X-ray diffraction peak position and thickness fringes.\cite{Herfort2004,jen07} In addition we measured Reflection high energy electron diffraction oscillations and X-ray oscillations of the layer-by-layer growth of Fe$_3$Si. We have used these methods because the flux rates are relatively low and cannot be determined directly with the sufficient accuracy. Two different types of MBE systems were used, one with relatively low growth rate\cite{jeni03} due to geometrical reasons, the other with higher growth rate.\cite{herfort03} Two nominal Fe$_3$Si film thicknesses were taken into account: 3~ML (before coalescence of growth islands, samples~1~and~2) and 6~ML (after coalescence of growth islands, samples~3~and~4). Some of the samples (samples~1~and~3) were capped with 4~nm of amorphous Ge deposited at T$_G$ = 150$^{\circ}$C, the remaining ones (samples~2~and~4) were characterized \textit{in--situ} immediately after the growth. Sample~5 contains a 36~nm thick Fe$_3$Si film on top of the GaAs(001) buffer layer. It was grown at high growth rate. The ordering of the thick Fe$_3$Si film near the Fe$_3$Si/GaAs interface is investigated using sample~5.

Synchrotron-based X-ray diffraction (XRD) was performed in grazing incidence geometry at the PHARAO U-125/2 KMC beamline of the storage ring BESSY II (Berlin). The photon energy was 10~keV, with an energy resolution of $\Delta{E}/E\sim$10$^{-4}$. The simulations of the crystal truncation rods were performed as in Ref.~\onlinecite{kag09}~, where the disorder parameters $\alpha$ and $\beta$ where taken into account.

The GaAs 222 and 002 reflections are quasiforbidden and in that way the corresponding Fe$_3$Si maxima are not disturbed by an intense substrate reflection. In this manner the disorder parameter $\beta$ can be determined with high sensitivity. In the present work we restrict ourselves to this parameter, because the amount of material is extremely small and $\alpha$ cannot be determined due to intense substrate contribution for odd H,K,L.

Cross-sectional TEM specimens were prepared by mechanical lapping and polishing, followed by Ar ion milling. A TEM JEOL~JEM2100~F operated at 200~kV was used for high-resolution (HR) imaging.  The probe-C$_s$-corrected JEOL ARM200 operated in the scanning TEM (STEM) mode at 200~kV  was utilized for atomically resolved high-angle annular dark field imaging. In addition, corresponding image contrast simulations using the program \textit{JEMS} allowed for a certain interpretation of the image contrast.\cite{stadelmann2016}

\section{Results}
Fe$_3$Si islands of  sample~1 imaged by HR TEM (Figure~\ref{islands}) are approximately 4~MLs high and 3~nm in lateral size. The real height of 4~MLs is larger than the nominal film thickness 3~MLs, and so the coverage of the GaAs surface becomes smaller than 1.

Figure~\ref{XRD} shows the L-scans of all the samples of the 20L and 22L crystal truncation rods measured by grazing incidence diffraction (GID)  using synchrotron radiation. From comparison with simulations we obtain for sample~1 an island height of 4~MLs and a poor ordering of the Fe$_3$Si islands (with $\beta$~=~0.48, see Table~\ref{tab:tab1}, cf. also~\cite{Jenichen05}). The 202 maximum is fundamental and not sensitive to disorder.\cite{Jenichen05} Therefore we determined the island height from comparison of the experimental curve and the simulation of this 20L crystal truncation rod and used the same height for the simulation of the 22L-measurement as well. Obviously for the 22L-measurement the fringe period is deviating from the calculated value indicating inhomogeneity of the ordering.

The 202 diffraction maxima of Fe$_3$Si and GaAs overlap without peak shift, i.e. the Fe$_3$Si islands exhibit nearly ideal stoichiometry.\cite{herfort03,Herfort2004} For sample~1 the 222 and 002 maxima have characteristic shapes with a strong reduction of the Fe$_3$Si layer maxima, which are evidence for chemical disorder in the Fe$_3$Si. The difference with respect to a fully ordered film becomes obvious from comparison with sample~2 (Figure~\ref{XRD}). The 202 peaks of both samples are rather similar, because they are not sensitive to disorder, whereas the 222 peaks of the Fe$_3$Si islands differ, because in sample~1 the Fe$_3$Si is disordered, resulting in a reduced Fe$_3$Si 222 peak intensity, and in sample~2 it is fully ordered with $\beta$~=~0.0 (see Table~\ref{tab:tab1}), and the Fe$_3$Si 222 peak exhibits full intensity. Sample~3 contains a nominally 6~ML thick film, and was grown at high growth rate. It exhibits strong disorder with $\beta$~=~0.45. For sample~3 the thickness of the disordered region again does not coincide with the full film thickness due to inhomogeneous ordering resulting in different interference period lengths for fundamental and superlattice maxima, i.e. a disagreement of measurement and simulation.  Sample~4 is a nominally (7$\pm$1)~ML thick film, and was grown at low growth rate, and exhibits perfect ordering with $\beta$~=~0.0 (see Table~\ref{tab:tab1}). The measured thickness for this sample coincides with the nominal thickness, i.e. the coverage now equals one. The film is continuous now, all islands are coalesced. This would be the starting phase of Fe$_3$Si homoepitaxy.

\begin{figure}[h]
\includegraphics[width=0.85\linewidth,angle=0, clip]{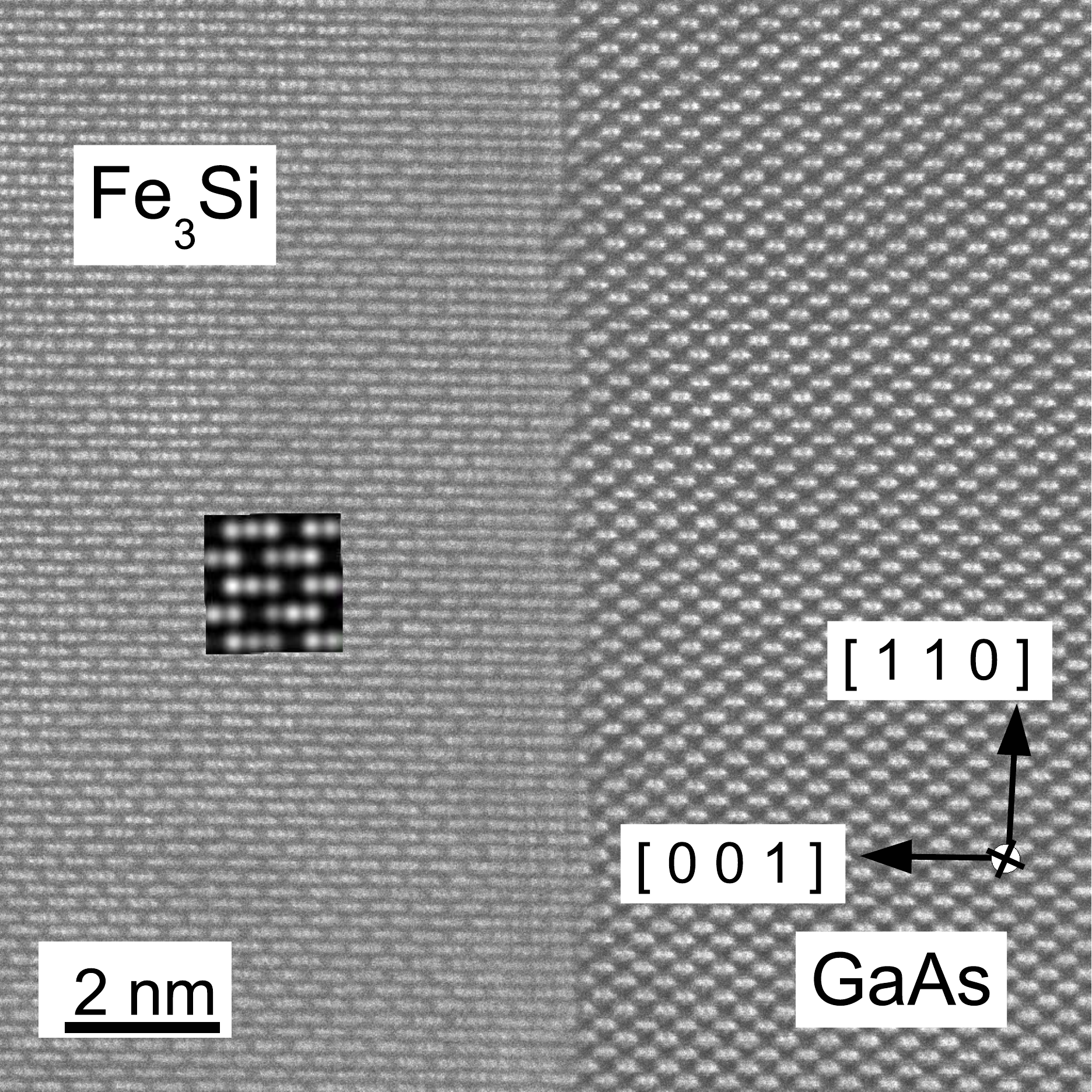}
\caption {Cross-section high-resolution scanning transmission electron micrograph of an Fe$_3$Si film epitaxially grown on the GaAs(001) substrate at a growth rate of 71~ML/h detected in the high-angle annular dark field mode. Near the Fe$_3$Si/GaAs interface we find evidence of disordering of the Fe$_3$Si. The inset shows a simulation for perfectly D0$_3$ ordered Fe$_3$Si.}
\label{TEM2}
\end{figure}

Figure~\ref{TEM2} shows the Fe$_3$Si/GaAs interface of sample~5 in a scanning transmission micrograph taken in the high-angle annular dark field (HAADF) mode of the STEM. The inset depicts a simulation for perfectly D0$_3$ ordered Fe$_3$Si with characteristic Fe-triples. In the z-contrast mode the Fe atoms give the highest scattering intensity, whereas the Si atoms scatter with lower intensity.\cite{Pennycook1990} The GaAs crystal structure is nearly ideal, however the Fe$_3$Si structure shows evidence of disordering near the interface, where the Fe-triples are blurred. The sample was grown at a growth rate of 71~ML/h. The disordering near the interface is connected to the disorder occuring during the starting phase of the epitaxial growth. We note, that a similar effect was obtained for the growth of lattice matched Co$_2$FeSi Heusler alloy film on GaAs(001).\cite{hashimoto2007jvst} The role of interdiffusion should be stronger for lower growth rate, however, perfect ordering is found for the lower growth rate. In this way the disorder found in our case is rather a growth phenomenon.

Neglecting the anisotropy of Fe$_3$Si we can make a rough estimate of the Fe$_3$Si/GaAs(001) interface energy $\gamma_{IF}$ using the simple formula from Ref.~\onlinecite{Gennes1985}

\begin{equation}
\gamma_{IF} = |\gamma_{GaAs}-\gamma_{Fe3Si}\cdot\cos(\theta)|,
\end{equation}
\label{surf}

where $\gamma_{IF}\approx100~meV/{\AA}^2$ taking into account a contact angle $\theta\approx33^{\circ}$ (Figure~\ref{islands}), the surface energy of GaAs $\gamma_{GaAs}\approx65~meV/{\AA}^2$, and the surface energy of Fe$_3$Si $\gamma_{Fe3Si}\approx200~meV/{\AA}^2$ (see above). The limited accuracy of this estimate does not allow for any conclusions about the influence of the growth rate on the interface energy.

\section{Conclusion}
Thin Fe$_3$Si islands and films grown on GaAs(001) exhibit long-range ordering which is depending on their growth rate. A sufficiently low enough growth rate can secure a fully ordered Fe$_3$Si lattice, whereas a higher growth rate leads to a nearly fully disordered film, which is however still lattice matched. The disorder occurring in the starting phase of the growth seems to be the reason for disorder observed near the Fe$_3$Si/GaAs interface. \\

\section{Acknowledgments}
The authors thank V.~Kaganer for helpful discussion and the simulation program for the X-ray data. We also thank H.~Kirmse and T.~Heil for the help with the HAADF micrograph shown in Figure~\ref{TEM2}. We are grateful C.~Herrmann and H.-P.~Sch\"{o}nherr for their support during the growth of the samples, and D.~Steffen for preparation of the TEM-samples. \\

\section{References}



\begin{thebibliography}{25}%
\makeatletter
\providecommand \@ifxundefined [1]{%
 \@ifx{#1\undefined}
}%
\providecommand \@ifnum [1]{%
 \ifnum #1\expandafter \@firstoftwo
 \else \expandafter \@secondoftwo
 \fi
}%
\providecommand \@ifx [1]{%
 \ifx #1\expandafter \@firstoftwo
 \else \expandafter \@secondoftwo
 \fi
}%
\providecommand \natexlab [1]{#1}%
\providecommand \enquote  [1]{``#1''}%
\providecommand \bibnamefont  [1]{#1}%
\providecommand \bibfnamefont [1]{#1}%
\providecommand \citenamefont [1]{#1}%
\providecommand \href@noop [0]{\@secondoftwo}%
\providecommand \href [0]{\begingroup \@sanitize@url \@href}%
\providecommand \@href[1]{\@@startlink{#1}\@@href}%
\providecommand \@@href[1]{\endgroup#1\@@endlink}%
\providecommand \@sanitize@url [0]{\catcode `\\12\catcode `\$12\catcode
  `\&12\catcode `\#12\catcode `\^12\catcode `\_12\catcode `\%12\relax}%
\providecommand \@@startlink[1]{}%
\providecommand \@@endlink[0]{}%
\providecommand \url  [0]{\begingroup\@sanitize@url \@url }%
\providecommand \@url [1]{\endgroup\@href {#1}{\urlprefix }}%
\providecommand \urlprefix  [0]{URL }%
\providecommand \Eprint [0]{\href }%
\providecommand \doibase [0]{http://dx.doi.org/}%
\providecommand \selectlanguage [0]{\@gobble}%
\providecommand \bibinfo  [0]{\@secondoftwo}%
\providecommand \bibfield  [0]{\@secondoftwo}%
\providecommand \translation [1]{[#1]}%
\providecommand \BibitemOpen [0]{}%
\providecommand \bibitemStop [0]{}%
\providecommand \bibitemNoStop [0]{.\EOS\space}%
\providecommand \EOS [0]{\spacefactor3000\relax}%
\providecommand \BibitemShut  [1]{\csname bibitem#1\endcsname}%
\let\auto@bib@innerbib\@empty
\bibitem [{\citenamefont {Volmer}\ and\ \citenamefont
  {Weber}(1926)}]{Volmer1926}%
  \BibitemOpen
  \bibfield  {author} {\bibinfo {author} {\bibfnamefont {M.}~\bibnamefont
  {Volmer}}\ and\ \bibinfo {author} {\bibfnamefont {A.}~\bibnamefont {Weber}},\
  }\href@noop {} {\bibfield  {journal} {\bibinfo  {journal} {Z. phys. Chem.}\
  }\textbf {\bibinfo {volume} {119}},\ \bibinfo {pages} {277} (\bibinfo {year}
  {1926})}\BibitemShut {NoStop}%
\bibitem [{\citenamefont {Raviswaran}\ \emph {et~al.}(2001)\citenamefont
  {Raviswaran}, \citenamefont {Liu}, \citenamefont {Kim}, \citenamefont
  {Cahill},\ and\ \citenamefont {Gibson}}]{Ravisvaran2001}%
  \BibitemOpen
  \bibfield  {author} {\bibinfo {author} {\bibfnamefont {A.}~\bibnamefont
  {Raviswaran}}, \bibinfo {author} {\bibfnamefont {C.-P.}\ \bibnamefont {Liu}},
  \bibinfo {author} {\bibfnamefont {J.}~\bibnamefont {Kim}}, \bibinfo {author}
  {\bibfnamefont {D.~G.}\ \bibnamefont {Cahill}}, \ and\ \bibinfo {author}
  {\bibfnamefont {G.}~\bibnamefont {Gibson}},\ }\href@noop {} {\bibfield
  {journal} {\bibinfo  {journal} {Phys. Rev. B}\ }\textbf {\bibinfo {volume}
  {63}},\ \bibinfo {pages} {125314} (\bibinfo {year} {2001})}\BibitemShut
  {NoStop}%
\bibitem [{\citenamefont {Kaganer}\ \emph {et~al.}(2009)\citenamefont
  {Kaganer}, \citenamefont {Jenichen}, \citenamefont {Shayduk}, \citenamefont
  {Braun},\ and\ \citenamefont {Riechert}}]{kag09}%
  \BibitemOpen
  \bibfield  {author} {\bibinfo {author} {\bibfnamefont {V.~M.}\ \bibnamefont
  {Kaganer}}, \bibinfo {author} {\bibfnamefont {B.}~\bibnamefont {Jenichen}},
  \bibinfo {author} {\bibfnamefont {R.}~\bibnamefont {Shayduk}}, \bibinfo
  {author} {\bibfnamefont {W.}~\bibnamefont {Braun}}, \ and\ \bibinfo {author}
  {\bibfnamefont {H.}~\bibnamefont {Riechert}},\ }\href@noop {} {\bibfield
  {journal} {\bibinfo  {journal} {Phys. Rev. Lett.}\ }\textbf {\bibinfo
  {volume} {102}},\ \bibinfo {pages} {016103} (\bibinfo {year}
  {2009})}\BibitemShut {NoStop}%
\bibitem [{\citenamefont {Placidi}\ \emph {et~al.}(2000)\citenamefont
  {Placidi}, \citenamefont {Fanfoni}, \citenamefont {Archiprete}, \citenamefont
  {Patella}, \citenamefont {Motta},\ and\ \citenamefont
  {Balzarotti}}]{Placidi2000}%
  \BibitemOpen
  \bibfield  {author} {\bibinfo {author} {\bibfnamefont {E.}~\bibnamefont
  {Placidi}}, \bibinfo {author} {\bibfnamefont {M.}~\bibnamefont {Fanfoni}},
  \bibinfo {author} {\bibfnamefont {F.}~\bibnamefont {Archiprete}}, \bibinfo
  {author} {\bibfnamefont {F.}~\bibnamefont {Patella}}, \bibinfo {author}
  {\bibfnamefont {N.}~\bibnamefont {Motta}}, \ and\ \bibinfo {author}
  {\bibfnamefont {A.}~\bibnamefont {Balzarotti}},\ }\href@noop {} {\bibfield
  {journal} {\bibinfo  {journal} {Mat. Sci. and Eng.}\ }\textbf {\bibinfo
  {volume} {B69-70}},\ \bibinfo {pages} {243} (\bibinfo {year}
  {2000})}\BibitemShut {NoStop}%
\bibitem [{\citenamefont {Moll}\ \emph {et~al.}(1996)\citenamefont {Moll},
  \citenamefont {Kley}, \citenamefont {Pehlke},\ and\ \citenamefont
  {Scheffler}}]{moll1996}%
  \BibitemOpen
  \bibfield  {author} {\bibinfo {author} {\bibfnamefont {N.}~\bibnamefont
  {Moll}}, \bibinfo {author} {\bibfnamefont {A.}~\bibnamefont {Kley}}, \bibinfo
  {author} {\bibfnamefont {E.}~\bibnamefont {Pehlke}}, \ and\ \bibinfo {author}
  {\bibfnamefont {M.}~\bibnamefont {Scheffler}},\ }\href@noop {} {\bibfield
  {journal} {\bibinfo  {journal} {Phys. Rev. B}\ }\textbf {\bibinfo {volume}
  {54}},\ \bibinfo {pages} {8844} (\bibinfo {year} {1996})}\BibitemShut
  {NoStop}%
\bibitem [{\citenamefont {Hafner}\ and\ \citenamefont
  {Spis\'{a}k}(2007)}]{Hafner2007}%
  \BibitemOpen
  \bibfield  {author} {\bibinfo {author} {\bibfnamefont {J.}~\bibnamefont
  {Hafner}}\ and\ \bibinfo {author} {\bibfnamefont {D.}~\bibnamefont
  {Spis\'{a}k}},\ }\href@noop {} {\bibfield  {journal} {\bibinfo  {journal}
  {Phys. Rev. B}\ }\textbf {\bibinfo {volume} {75}},\ \bibinfo {pages} {195411}
  (\bibinfo {year} {2007})}\BibitemShut {NoStop}%
\bibitem [{\citenamefont {Herfort}\ \emph {et~al.}(2003)\citenamefont
  {Herfort}, \citenamefont {Sch\"onherr},\ and\ \citenamefont
  {Ploog}}]{herfort03}%
  \BibitemOpen
  \bibfield  {author} {\bibinfo {author} {\bibfnamefont {J.}~\bibnamefont
  {Herfort}}, \bibinfo {author} {\bibfnamefont {H.-P.}\ \bibnamefont
  {Sch\"onherr}}, \ and\ \bibinfo {author} {\bibfnamefont {K.~H.}\ \bibnamefont
  {Ploog}},\ }\href@noop {} {\bibfield  {journal} {\bibinfo  {journal} {Appl.
  Phys. Lett.}\ }\textbf {\bibinfo {volume} {83}},\ \bibinfo {pages} {3912}
  (\bibinfo {year} {2003})}\BibitemShut {NoStop}%
\bibitem [{\citenamefont {Herfort}\ \emph {et~al.}(2004)\citenamefont
  {Herfort}, \citenamefont {Schoenherr}, \citenamefont {Friedland},\ and\
  \citenamefont {Ploog}}]{Herfort2004}%
  \BibitemOpen
  \bibfield  {author} {\bibinfo {author} {\bibfnamefont {J.}~\bibnamefont
  {Herfort}}, \bibinfo {author} {\bibfnamefont {H.~P.}\ \bibnamefont
  {Schoenherr}}, \bibinfo {author} {\bibfnamefont {K.~J.}\ \bibnamefont
  {Friedland}}, \ and\ \bibinfo {author} {\bibfnamefont {K.~H.}\ \bibnamefont
  {Ploog}},\ }\href@noop {} {\bibfield  {journal} {\bibinfo  {journal} {J. Vac.
  Sci. Technol. B}\ }\textbf {\bibinfo {volume} {22}},\ \bibinfo {pages} {2073}
  (\bibinfo {year} {2004})}\BibitemShut {NoStop}%
\bibitem [{\citenamefont {Jenichen}\ \emph {et~al.}(2005)\citenamefont
  {Jenichen}, \citenamefont {Kaganer}, \citenamefont {Herfort}, \citenamefont
  {Satapathy}, \citenamefont {Sch\"onherr}, \citenamefont {Braun},\ and\
  \citenamefont {Ploog}}]{Jenichen05}%
  \BibitemOpen
  \bibfield  {author} {\bibinfo {author} {\bibfnamefont {B.}~\bibnamefont
  {Jenichen}}, \bibinfo {author} {\bibfnamefont {V.~M.}\ \bibnamefont
  {Kaganer}}, \bibinfo {author} {\bibfnamefont {J.}~\bibnamefont {Herfort}},
  \bibinfo {author} {\bibfnamefont {D.~K.}\ \bibnamefont {Satapathy}}, \bibinfo
  {author} {\bibfnamefont {H.~P.}\ \bibnamefont {Sch\"onherr}}, \bibinfo
  {author} {\bibfnamefont {W.}~\bibnamefont {Braun}}, \ and\ \bibinfo {author}
  {\bibfnamefont {K.~H.}\ \bibnamefont {Ploog}},\ }\href@noop {} {\bibfield
  {journal} {\bibinfo  {journal} {Phys. Rev. B}\ }\textbf {\bibinfo {volume}
  {72}},\ \bibinfo {pages} {075329} (\bibinfo {year} {2005})}\BibitemShut
  {NoStop}%
\bibitem [{\citenamefont {Niculescu}\ \emph {et~al.}(1976)\citenamefont
  {Niculescu}, \citenamefont {Raj}, \citenamefont {Budnick}, \citenamefont
  {Burch}, \citenamefont {Hines},\ and\ \citenamefont {Menotti}}]{niculescu76}%
  \BibitemOpen
  \bibfield  {author} {\bibinfo {author} {\bibfnamefont {V.}~\bibnamefont
  {Niculescu}}, \bibinfo {author} {\bibfnamefont {K.}~\bibnamefont {Raj}},
  \bibinfo {author} {\bibfnamefont {J.~I.}\ \bibnamefont {Budnick}}, \bibinfo
  {author} {\bibfnamefont {T.~J.}\ \bibnamefont {Burch}}, \bibinfo {author}
  {\bibfnamefont {W.~A.}\ \bibnamefont {Hines}}, \ and\ \bibinfo {author}
  {\bibfnamefont {A.~H.}\ \bibnamefont {Menotti}},\ }\href@noop {} {\bibfield
  {journal} {\bibinfo  {journal} {Phys. Rev. B}\ }\textbf {\bibinfo {volume}
  {14}},\ \bibinfo {pages} {4160} (\bibinfo {year} {1976})}\BibitemShut
  {NoStop}%
\bibitem [{\citenamefont {Hongzhi}\ \emph {et~al.}(2007)\citenamefont
  {Hongzhi}, \citenamefont {Zhiyong}, \citenamefont {Li}, \citenamefont
  {Shifeng}, \citenamefont {Heyan}, \citenamefont {Jingping}, \citenamefont
  {Yangxian},\ and\ \citenamefont {Guangheng}}]{Hongzhi2007}%
  \BibitemOpen
  \bibfield  {author} {\bibinfo {author} {\bibfnamefont {L.}~\bibnamefont
  {Hongzhi}}, \bibinfo {author} {\bibfnamefont {Z.}~\bibnamefont {Zhiyong}},
  \bibinfo {author} {\bibfnamefont {M.}~\bibnamefont {Li}}, \bibinfo {author}
  {\bibfnamefont {X.}~\bibnamefont {Shifeng}}, \bibinfo {author} {\bibfnamefont
  {L.}~\bibnamefont {Heyan}}, \bibinfo {author} {\bibfnamefont
  {Q.}~\bibnamefont {Jingping}}, \bibinfo {author} {\bibfnamefont
  {L.}~\bibnamefont {Yangxian}}, \ and\ \bibinfo {author} {\bibfnamefont
  {W.}~\bibnamefont {Guangheng}},\ }\href@noop {} {\bibfield  {journal}
  {\bibinfo  {journal} {J. Phys. D: Appl. Phys.}\ }\textbf {\bibinfo {volume}
  {40}},\ \bibinfo {pages} {7121} (\bibinfo {year} {2007})}\BibitemShut
  {NoStop}%
\bibitem [{\citenamefont {Noor}\ \emph {et~al.}(2013)\citenamefont {Noor},
  \citenamefont {Barsukov}, \citenamefont {Ozkan}, \citenamefont {Elbers},
  \citenamefont {Melnichak}, \citenamefont {Lindner}, \citenamefont {Farle},\
  and\ \citenamefont {Koehler}}]{noor2013}%
  \BibitemOpen
  \bibfield  {author} {\bibinfo {author} {\bibfnamefont {S.}~\bibnamefont
  {Noor}}, \bibinfo {author} {\bibfnamefont {I.}~\bibnamefont {Barsukov}},
  \bibinfo {author} {\bibfnamefont {M.~S.}\ \bibnamefont {Ozkan}}, \bibinfo
  {author} {\bibfnamefont {L.}~\bibnamefont {Elbers}}, \bibinfo {author}
  {\bibfnamefont {N.}~\bibnamefont {Melnichak}}, \bibinfo {author}
  {\bibfnamefont {J.}~\bibnamefont {Lindner}}, \bibinfo {author} {\bibfnamefont
  {M.}~\bibnamefont {Farle}}, \ and\ \bibinfo {author} {\bibfnamefont
  {U.}~\bibnamefont {Koehler}},\ }\href@noop {} {\bibfield  {journal} {\bibinfo
   {journal} {J. Appl. Phys.}\ }\textbf {\bibinfo {volume} {113}},\ \bibinfo
  {pages} {103908} (\bibinfo {year} {2013})}\BibitemShut {NoStop}%
\bibitem [{\citenamefont {Jenichen}\ \emph {et~al.}(2007)\citenamefont
  {Jenichen}, \citenamefont {Kaganer}, \citenamefont {Braun}, \citenamefont
  {Herfort}, \citenamefont {Shayduk},\ and\ \citenamefont {Ploog}}]{jen07}%
  \BibitemOpen
  \bibfield  {author} {\bibinfo {author} {\bibfnamefont {B.}~\bibnamefont
  {Jenichen}}, \bibinfo {author} {\bibfnamefont {V.~M.}\ \bibnamefont
  {Kaganer}}, \bibinfo {author} {\bibfnamefont {W.}~\bibnamefont {Braun}},
  \bibinfo {author} {\bibfnamefont {J.}~\bibnamefont {Herfort}}, \bibinfo
  {author} {\bibfnamefont {R.}~\bibnamefont {Shayduk}}, \ and\ \bibinfo
  {author} {\bibfnamefont {K.~H.}\ \bibnamefont {Ploog}},\ }\href@noop {}
  {\bibfield  {journal} {\bibinfo  {journal} {Thin Solid Films}\ }\textbf
  {\bibinfo {volume} {515}},\ \bibinfo {pages} {5611} (\bibinfo {year}
  {2007})}\BibitemShut {NoStop}%
\bibitem [{\citenamefont {Jenichen}\ \emph {et~al.}(2008)\citenamefont
  {Jenichen}, \citenamefont {Kaganer}, \citenamefont {Braun}, \citenamefont
  {Shayduk}, \citenamefont {Tinkham},\ and\ \citenamefont {Herfort}}]{jen08a}%
  \BibitemOpen
  \bibfield  {author} {\bibinfo {author} {\bibfnamefont {B.}~\bibnamefont
  {Jenichen}}, \bibinfo {author} {\bibfnamefont {V.~M.}\ \bibnamefont
  {Kaganer}}, \bibinfo {author} {\bibfnamefont {W.}~\bibnamefont {Braun}},
  \bibinfo {author} {\bibfnamefont {R.}~\bibnamefont {Shayduk}}, \bibinfo
  {author} {\bibfnamefont {B.~P.}\ \bibnamefont {Tinkham}}, \ and\ \bibinfo
  {author} {\bibfnamefont {J.}~\bibnamefont {Herfort}},\ }\href@noop {}
  {\bibfield  {journal} {\bibinfo  {journal} {J. Mat. Sci.: Mater. Electron.}\
  }\textbf {\bibinfo {volume} {19}},\ \bibinfo {pages} {199} (\bibinfo {year}
  {2008})}\BibitemShut {NoStop}%
\bibitem [{\citenamefont {Ionescu}\ \emph {et~al.}(2005)\citenamefont
  {Ionescu}, \citenamefont {Waz}, \citenamefont {Trypiniotis}, \citenamefont
  {G\"urtler}, \citenamefont {Garcia-Miquel}, \citenamefont {Bland},
  \citenamefont {Vickers}, \citenamefont {Dalgliesh}, \citenamefont
  {Langridge}, \citenamefont {Bugoslavsky}, \citenamefont {Miyoshi},
  \citenamefont {cohen},\ and\ \citenamefont {Ziebeck}}]{ionescu05}%
  \BibitemOpen
  \bibfield  {author} {\bibinfo {author} {\bibfnamefont {A.}~\bibnamefont
  {Ionescu}}, \bibinfo {author} {\bibfnamefont {C.~A.~F.}\ \bibnamefont {Waz}},
  \bibinfo {author} {\bibfnamefont {T.}~\bibnamefont {Trypiniotis}}, \bibinfo
  {author} {\bibfnamefont {C.~M.}\ \bibnamefont {G\"urtler}}, \bibinfo {author}
  {\bibfnamefont {H.}~\bibnamefont {Garcia-Miquel}}, \bibinfo {author}
  {\bibfnamefont {J.~A.~C.}\ \bibnamefont {Bland}}, \bibinfo {author}
  {\bibfnamefont {M.~E.}\ \bibnamefont {Vickers}}, \bibinfo {author}
  {\bibfnamefont {R.~M.}\ \bibnamefont {Dalgliesh}}, \bibinfo {author}
  {\bibfnamefont {C.}~\bibnamefont {Langridge}}, \bibinfo {author}
  {\bibfnamefont {Y.}~\bibnamefont {Bugoslavsky}}, \bibinfo {author}
  {\bibfnamefont {Y.}~\bibnamefont {Miyoshi}}, \bibinfo {author} {\bibfnamefont
  {L.~F.}\ \bibnamefont {cohen}}, \ and\ \bibinfo {author} {\bibfnamefont
  {K.~R.~A.}\ \bibnamefont {Ziebeck}},\ }\href@noop {} {\bibfield  {journal}
  {\bibinfo  {journal} {Phys. Rev. B}\ }\textbf {\bibinfo {volume} {71}},\
  \bibinfo {pages} {094401} (\bibinfo {year} {2005})}\BibitemShut {NoStop}%
\bibitem [{\citenamefont {Herfort}\ \emph {et~al.}(2005)\citenamefont
  {Herfort}, \citenamefont {Sch\"onherr}, \citenamefont {Kawaharazuka},
  \citenamefont {Ramsteiner},\ and\ \citenamefont {Ploog}}]{Herfort2005}%
  \BibitemOpen
  \bibfield  {author} {\bibinfo {author} {\bibfnamefont {J.}~\bibnamefont
  {Herfort}}, \bibinfo {author} {\bibfnamefont {H.-P.}\ \bibnamefont
  {Sch\"onherr}}, \bibinfo {author} {\bibfnamefont {A.}~\bibnamefont
  {Kawaharazuka}}, \bibinfo {author} {\bibfnamefont {M.}~\bibnamefont
  {Ramsteiner}}, \ and\ \bibinfo {author} {\bibfnamefont {K.~H.}\ \bibnamefont
  {Ploog}},\ }\href@noop {} {\bibfield  {journal} {\bibinfo  {journal} {J.
  Cryst. Growth}\ }\textbf {\bibinfo {volume} {278}},\ \bibinfo {pages} {666}
  (\bibinfo {year} {2005})}\BibitemShut {NoStop}%
\bibitem [{\citenamefont {Gusenbauer}\ \emph {et~al.}(2011)\citenamefont
  {Gusenbauer}, \citenamefont {Ashraf}, \citenamefont {Stangl}, \citenamefont
  {Hesser}, \citenamefont {Plach}, \citenamefont {Meingast}, \citenamefont
  {Kothleitner},\ and\ \citenamefont {Koch}}]{Gusenbauer2011}%
  \BibitemOpen
  \bibfield  {author} {\bibinfo {author} {\bibfnamefont {C.}~\bibnamefont
  {Gusenbauer}}, \bibinfo {author} {\bibfnamefont {T.}~\bibnamefont {Ashraf}},
  \bibinfo {author} {\bibfnamefont {J.}~\bibnamefont {Stangl}}, \bibinfo
  {author} {\bibfnamefont {G.}~\bibnamefont {Hesser}}, \bibinfo {author}
  {\bibfnamefont {T.}~\bibnamefont {Plach}}, \bibinfo {author} {\bibfnamefont
  {A.}~\bibnamefont {Meingast}}, \bibinfo {author} {\bibfnamefont
  {G.}~\bibnamefont {Kothleitner}}, \ and\ \bibinfo {author} {\bibfnamefont
  {R.}~\bibnamefont {Koch}},\ }\href@noop {} {\bibfield  {journal} {\bibinfo
  {journal} {Phys. Rev. B}\ }\textbf {\bibinfo {volume} {83}},\ \bibinfo
  {pages} {035319} (\bibinfo {year} {2011})}\BibitemShut {NoStop}%
\bibitem [{\citenamefont {Krumme}\ \emph {et~al.}(2009)\citenamefont {Krumme},
  \citenamefont {Weis}, \citenamefont {Herper}, \citenamefont {Stromberg},
  \citenamefont {Antoniak}, \citenamefont {Warland}, \citenamefont {Schuster},
  \citenamefont {Srivastava}, \citenamefont {Walterfang}, \citenamefont
  {Fauth}, \citenamefont {Min\'{a}r}, \citenamefont {Ebert}, \citenamefont
  {Entel}, \citenamefont {Keune},\ and\ \citenamefont {Wende}}]{Krumme2009}%
  \BibitemOpen
  \bibfield  {author} {\bibinfo {author} {\bibfnamefont {B.}~\bibnamefont
  {Krumme}}, \bibinfo {author} {\bibfnamefont {C.}~\bibnamefont {Weis}},
  \bibinfo {author} {\bibfnamefont {H.~C.}\ \bibnamefont {Herper}}, \bibinfo
  {author} {\bibfnamefont {F.}~\bibnamefont {Stromberg}}, \bibinfo {author}
  {\bibfnamefont {C.}~\bibnamefont {Antoniak}}, \bibinfo {author}
  {\bibfnamefont {A.}~\bibnamefont {Warland}}, \bibinfo {author} {\bibfnamefont
  {E.}~\bibnamefont {Schuster}}, \bibinfo {author} {\bibfnamefont
  {P.}~\bibnamefont {Srivastava}}, \bibinfo {author} {\bibfnamefont
  {M.}~\bibnamefont {Walterfang}}, \bibinfo {author} {\bibfnamefont
  {K.}~\bibnamefont {Fauth}}, \bibinfo {author} {\bibfnamefont
  {J.}~\bibnamefont {Min\'{a}r}}, \bibinfo {author} {\bibfnamefont
  {H.}~\bibnamefont {Ebert}}, \bibinfo {author} {\bibfnamefont
  {P.}~\bibnamefont {Entel}}, \bibinfo {author} {\bibfnamefont
  {W.}~\bibnamefont {Keune}}, \ and\ \bibinfo {author} {\bibfnamefont
  {H.}~\bibnamefont {Wende}},\ }\href@noop {} {\bibfield  {journal} {\bibinfo
  {journal} {Phys. Rev. B}\ }\textbf {\bibinfo {volume} {80}},\ \bibinfo
  {pages} {144403} (\bibinfo {year} {2009})}\BibitemShut {NoStop}%
\bibitem [{\citenamefont {Liou}\ \emph {et~al.}(1993)\citenamefont {Liou},
  \citenamefont {Malhotra}, \citenamefont {Shen}, \citenamefont {Hong},
  \citenamefont {Kwo}, \citenamefont {Chen},\ and\ \citenamefont
  {Mannaerts}}]{liou93}%
  \BibitemOpen
  \bibfield  {author} {\bibinfo {author} {\bibfnamefont {S.~H.}\ \bibnamefont
  {Liou}}, \bibinfo {author} {\bibfnamefont {S.~S.}\ \bibnamefont {Malhotra}},
  \bibinfo {author} {\bibfnamefont {J.~X.}\ \bibnamefont {Shen}}, \bibinfo
  {author} {\bibfnamefont {M.}~\bibnamefont {Hong}}, \bibinfo {author}
  {\bibfnamefont {J.}~\bibnamefont {Kwo}}, \bibinfo {author} {\bibfnamefont
  {H.~S.}\ \bibnamefont {Chen}}, \ and\ \bibinfo {author} {\bibfnamefont
  {J.~P.}\ \bibnamefont {Mannaerts}},\ }\href@noop {} {\bibfield  {journal}
  {\bibinfo  {journal} {J. Appl. Phys.}\ }\textbf {\bibinfo {volume} {73}},\
  \bibinfo {pages} {6766} (\bibinfo {year} {1993})}\BibitemShut {NoStop}%
\bibitem [{\citenamefont {Herfort}\ \emph {et~al.}(2008)\citenamefont
  {Herfort}, \citenamefont {Sch\"onherr},\ and\ \citenamefont
  {Jenichen}}]{Herfort2008}%
  \BibitemOpen
  \bibfield  {author} {\bibinfo {author} {\bibfnamefont {J.}~\bibnamefont
  {Herfort}}, \bibinfo {author} {\bibfnamefont {H.-P.}\ \bibnamefont
  {Sch\"onherr}}, \ and\ \bibinfo {author} {\bibfnamefont {B.}~\bibnamefont
  {Jenichen}},\ }\href@noop {} {\bibfield  {journal} {\bibinfo  {journal} {J.
  Appl. Phys.}\ }\textbf {\bibinfo {volume} {103}},\ \bibinfo {pages} {07B506}
  (\bibinfo {year} {2008})}\BibitemShut {NoStop}%
\bibitem [{\citenamefont {Jenichen}\ \emph {et~al.}(2003)\citenamefont
  {Jenichen}, \citenamefont {Braun}, \citenamefont {Kaganer}, \citenamefont
  {Shtukenberg}, \citenamefont {D\"aweritz}, \citenamefont {Schulz},\ and\
  \citenamefont {Ploog}}]{jeni03}%
  \BibitemOpen
  \bibfield  {author} {\bibinfo {author} {\bibfnamefont {B.}~\bibnamefont
  {Jenichen}}, \bibinfo {author} {\bibfnamefont {W.}~\bibnamefont {Braun}},
  \bibinfo {author} {\bibfnamefont {V.~M.}\ \bibnamefont {Kaganer}}, \bibinfo
  {author} {\bibfnamefont {A.~G.}\ \bibnamefont {Shtukenberg}}, \bibinfo
  {author} {\bibfnamefont {L.}~\bibnamefont {D\"aweritz}}, \bibinfo {author}
  {\bibfnamefont {C.~G.}\ \bibnamefont {Schulz}}, \ and\ \bibinfo {author}
  {\bibfnamefont {K.~H.}\ \bibnamefont {Ploog}},\ }\href@noop {} {\bibfield
  {journal} {\bibinfo  {journal} {Rev. Sci. Instr.}\ }\textbf {\bibinfo
  {volume} {74}},\ \bibinfo {pages} {1267} (\bibinfo {year}
  {2003})}\BibitemShut {NoStop}%
\bibitem [{\citenamefont {Stadelmann}(2016)}]{stadelmann2016}%
  \BibitemOpen
  \bibfield  {author} {\bibinfo {author} {\bibfnamefont {P.}~\bibnamefont
  {Stadelmann}},\ }\href@noop {} {\bibfield  {journal} {\bibinfo  {journal}
  {Electron Microscopy Simulation progam \textit{JEMS}, version 4.xx,
  http://www.jems-saas.ch/, Lausanne,}\ } (\bibinfo {year} {2016})}\BibitemShut
  {NoStop}%
\bibitem [{\citenamefont {Pennycook}\ and\ \citenamefont
  {Jesson}(1990)}]{Pennycook1990}%
  \BibitemOpen
  \bibfield  {author} {\bibinfo {author} {\bibfnamefont {S.~J.}\ \bibnamefont
  {Pennycook}}\ and\ \bibinfo {author} {\bibfnamefont {D.~E.}\ \bibnamefont
  {Jesson}},\ }\href@noop {} {\bibfield  {journal} {\bibinfo  {journal} {Phys.
  Rev. Lett.}\ }\textbf {\bibinfo {volume} {64}},\ \bibinfo {pages} {938}
  (\bibinfo {year} {1990})}\BibitemShut {NoStop}%
\bibitem [{\citenamefont {Hashimoto}\ \emph {et~al.}(2007)\citenamefont
  {Hashimoto}, \citenamefont {Trampert}, \citenamefont {Herfort},\ and\
  \citenamefont {Ploog}}]{hashimoto2007jvst}%
  \BibitemOpen
  \bibfield  {author} {\bibinfo {author} {\bibfnamefont {M.}~\bibnamefont
  {Hashimoto}}, \bibinfo {author} {\bibfnamefont {A.}~\bibnamefont {Trampert}},
  \bibinfo {author} {\bibfnamefont {J.}~\bibnamefont {Herfort}}, \ and\
  \bibinfo {author} {\bibfnamefont {K.~H.}\ \bibnamefont {Ploog}},\ }\href@noop
  {} {\bibfield  {journal} {\bibinfo  {journal} {J. Vac. Sci. Technol. B}\
  }\textbf {\bibinfo {volume} {25}},\ \bibinfo {pages} {1453} (\bibinfo {year}
  {2007})}\BibitemShut {NoStop}%
\bibitem [{\citenamefont {de~Gennes}(1985)}]{Gennes1985}%
  \BibitemOpen
  \bibfield  {author} {\bibinfo {author} {\bibfnamefont {P.~G.}\ \bibnamefont
  {de~Gennes}},\ }\href@noop {} {\bibfield  {journal} {\bibinfo  {journal}
  {Rev. of Mod. Phys.}\ }\textbf {\bibinfo {volume} {57}},\ \bibinfo {pages}
  {827} (\bibinfo {year} {1985})}\BibitemShut {NoStop}%
\end{thebibliography}
%

\end{document}